\documentclass{appolb}
\usepackage{epsfig}

\usepackage{amsmath}
\usepackage{hhline}
\usepackage{amssymb}
\usepackage{times}
\usepackage{cite}
\usepackage{array}
\usepackage{multirow}

\begin{document}
\title{Recent Results in Prompt Photon Production%
\thanks{Invited talk at Photon2005, 31st - 4th September, Warsaw}%
}
\author{J\"org Gayler
\address{DESY}
}
\maketitle
\begin{abstract}
An introduction is given to recent results in prompt photon production in 
different reactions.
\end{abstract}
\PACS{PACS numbers come here}
  
\section{Introduction}
    Light  played always a key role in the attempts to understand early states
    of hadronic matter.
    In early reports on the creation of the world~\cite{Moses},
    light provided clarity and structure. Nowadays, we can see 
    the universe back to some $10^5$ years after the big bang by observation
    of light. In the microscopic world
    photons tell us about the original hard interactions through
    fire balls created in  nucleus-nucleus collisions. 
    Photons also give a rather clear message on partonic patterns, 
    in contrast to
    quarks and gluons which are not directly observable.
    Only the last two point will be further discussed in this report.
    
\begin{figure}[htb] \unitlength 1cm
   \begin{center}
   \begin{picture}(6.,1.8)
    \put(-2.4,-0.1){\includegraphics[width=0.85\textwidth]{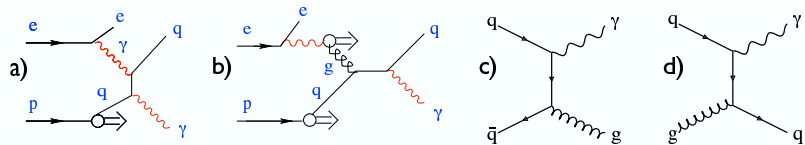}}
   \end{picture}
   \caption{Examples of LO graphs for prompt photon production in $\gamma p$
     (a,b) and hadron-hadron interactions (c,d).}
   \label{fig:graphs}
   \end{center}
   \end{figure}

    Usually photons are called "prompt'' (or "direct''), if they are coupling to 
    interacting partons, in contrast to photons from hadron decays
    or photons emitted by leptons. Figs.~\ref{fig:graphs} shows examples of
    leading order (LO) graphs of prompt
    photon emissions in $ep$ and hadron-hadron interactions.
 The $ep$ interactions in Figs.~\ref{fig:graphs}a) and b) are called
 photoproduction, if the photon virtuality $Q^2$ is small,
  (typically $< 1$ GeV$^2$),
  which means
 in case of the HERA experiments~\cite{Breitweg:1999su,Aktas:2004uv}
  that the scattered electron stays in the beam pipe.
 The photon can interact directly (Fig.~\ref{fig:graphs}a) or fluctuate
 into a hadronic state, part of which interacts with the incident proton
 (Fig.~\ref{fig:graphs}b).

 There is substantial interest in the observation of prompt photons
 as they are
    more directly related to partonic interactions than jets and 
   sensitive to the gluon content of the interacting particles
   (resolved photon and proton, Figs.~\ref{fig:graphs}b) and d) respectively).
They are an
  important background at searches at the LHC (e.g. $Higgs \rightarrow \gamma \gamma $),
   and, as they are not strongly interacting, they help to disentangle
   in nucleus-nucleus collisions
    effects of initial or final
   state interactions, of a quark gluon plasma or hadron gas...

 
   Various calculations exist in next to leading order (NLO) perturbative QCD (pQCD) based on next to leading order (NLO) matrix elements 
 and, in most cases, collinear parton densities (pdfs)
 of the interacting particles.
 In recent analyses also $k_{\rm t}$ factorised pdfs have been used~\cite{Lipatov:2005wk}
 for $\gamma p$ and $pp$ interactions.
  Further non-perturbative
 elements enter the calculations.
 Besides the $\gamma$'s indicated in Fig.~\ref{fig:graphs}, there are  $\gamma$'s from fragmentation processes of quarks
 and gluons which are part of the calculated signal.
  Detailed comparisons
 with experimental data require also simulation of the hadronic final
 state. First, because photons may be measured together with jets instead of inclusively,
 and second, because some  experiments require
 an isolation cone for the measured prompt photons. 
   
 Isolation of the prompt photon candidates is required in
 many  experiments to cope
 with the large background of photons from $\pi^0$ 
 and $\eta$ decay which may not be resolved as single $\gamma$'s
 in calorimetric measurements. The prompt photon signal is then determined
 by sophisticated shower shape analyses.
  Other experiments work without an explicit isolation
 condition
 and subtract measured  $\pi^0$ and $\eta$ yields.~\footnote{from the experimental data mentioned in this report,
  belong refs~\cite{Breitweg:1999su,Aktas:2004uv,Abbiendi:2003kf,Acosta:2002ya,D0:2005} 
  to the first and refs~\cite{Apanasevich:2004dr,Aggarwal:2000th,Adler:2005ig} to the second group.}

 In the following a few recent results
 of experiments with characteristics given in Table~1 will be 
  shortly discussed. 

\section{Prompt Photons in $\gamma p$ at HERA}

Recent results from H1 on inclusive prompt photons
 show that
 NLO calculations~\cite{Fontannaz:2001ek,Krawczyk:2001tz}
 describe
the measured distributions well in shape, being however low by 
about 30\% in normalisation,
  when
corrections for hadronisation are applied using the leading
order plus parton shower Monte Carlo (MC) programs of PYTHIA and HERWIG.
The MC generators themselves are also low by a similar amount.
A similar discrepancy was observed
in $\gamma \gamma $ interactions
by OPAL~\cite{Abbiendi:2003kf}.
If a jet is  required in addition to the prompt $\gamma$,
 the NLO description is good in various
distributions (see the figures in refs~\cite{Aktas:2004uv,Janssen:2005yz}).
One may speculate, that here more LO like configurations are selected which may reduce the 
phase space for higher order emissions.
  See~\cite{Chekanov:2004wr,Janssen:2005yz} for first results in DIS.

\section{Prompt Photons in Hadronic Reactions}

\begin{figure}[htb] \unitlength 1cm
   \begin{center}
   \begin{picture}(6,6.0)
   \put(-3.3,-0.2){\includegraphics[width=0.95\textwidth]{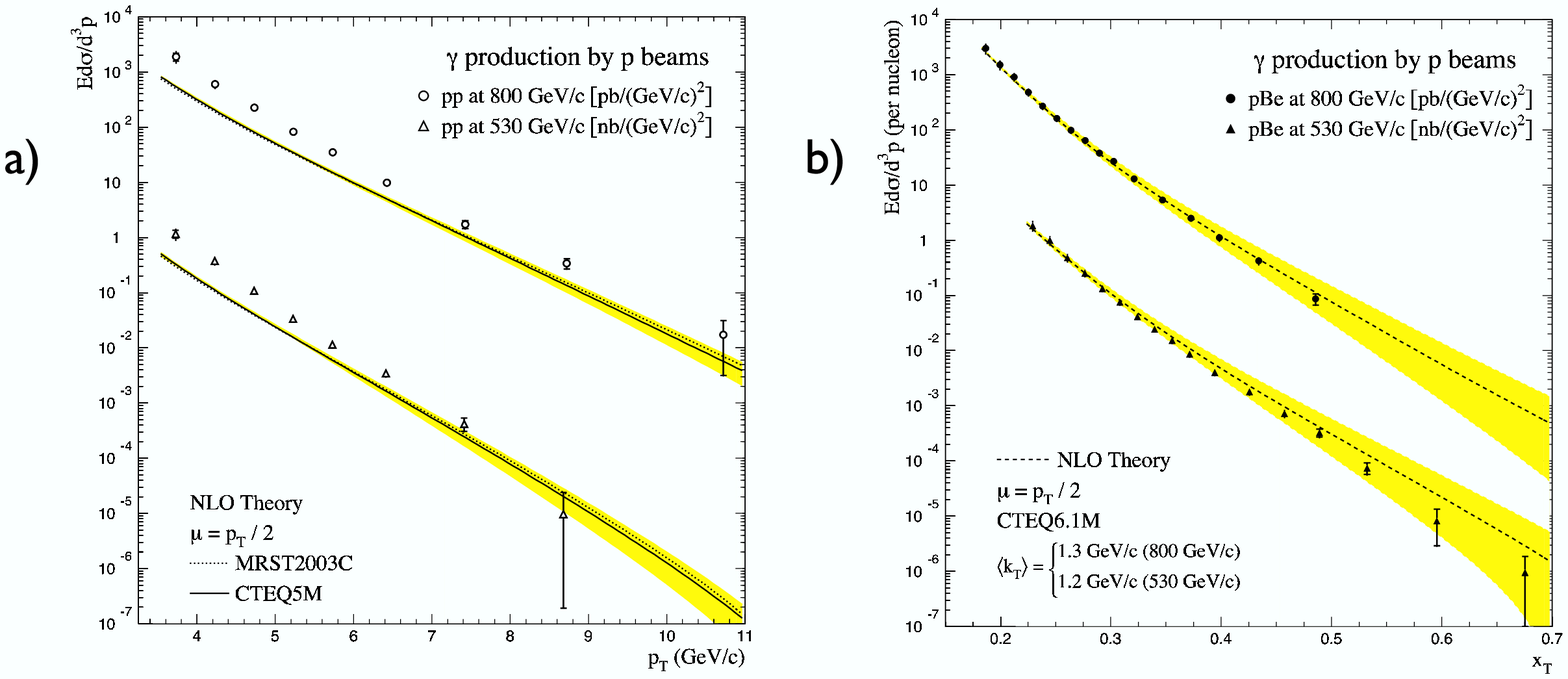}} 
   \end{picture}
   \caption{E706 results  compared with pQCD.
    a) $pp \rightarrow \gamma X$ vs. $p_{\rm t}$ , b) $pBe \rightarrow \gamma X$
     vs. $x_{\rm t} = 2 p_{\rm t}/\sqrt{s}$
     with $k_{\rm t}$~smearing.}
   \label{fig:pE706}
   \end{center}
   \end{figure}

  Notoriously, there are difficulties to describe prompt photon production
  in pQCD,
  particularly at fixed target energies. For example the high statistics data of
the E706 collaboration~\cite{Apanasevich:2004dr} (see Fig.~\ref{fig:pE706})
  at $\sqrt{s} = 32$ and 39 GeV are above NLO theory~\cite{Aurenche:1983ws} by about a factor 2 at low $p_{\rm t}$.
   Agreement is reached by an ad hoc smearing by an intrinsic
  parton $k_{\rm t}$ of the protons $\gtrsim 1$ GeV (see e.g.~\cite{deFlorian:2005wf}
  for theoretical improvements by
  resummations).   
  The deviations are smaller at high energies, 
  but the CDF data~\cite{Acosta:2002ya} at $\sqrt{s} = 1.8$ TeV
  show also steeper a $p_{\rm t}$ dependence than predicted~\cite{Gluck:1994iz}.
  It is interesting to note
  that more recent CDF results~\cite{Acosta:2004bg}
   which are based on photon conversions
  are consistent with the former calorimetric~\cite{Acosta:2002ya} measurements
  with quite different systematics.
  However, the preliminary D0 data~\cite{D0:2005,Soldner} from
   Tevatron Run 2 at $\sqrt{s} = 1.96$~TeV
  are consistent with NLO theory~\cite{Catani:2002ny} within errors.
  See ~\cite{Soldner} for di-photon results from CDF.
  
\begin{figure}[htb] \unitlength 1cm
   \begin{center}
   \begin{picture}(6,5.)
   \put(-3.,-0.1){\includegraphics[width=0.95\textwidth]{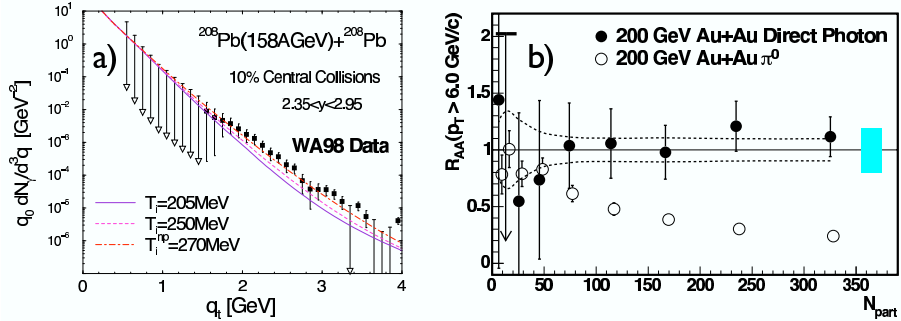}}
   \end{picture}
   \caption{a) WA98 results described by QGP effects and pQCD; b) PHENIX  
   yields for 
   $\gamma$ and $\pi^0$ at $p_{\rm t} > 6$ GeV 
    vs. the number of participating nucleons,
    scaled from AuAu to NN.
    }
   \label{fig:nucl}
   \end{center}
   \end{figure}

  In nucleus-nucleus interactions, thermal photons are expected due to 
  quark-gluon-plasma (QGP) or, at even smaller $p_{\rm t}$, from a hadron gas.
  Fig.~\ref{fig:nucl}a shows the interpretation of the WA98 Pb-Pb data
  ($\sqrt{s_{NN}} = 17.3$ GeV)
  in terms of a convolution~\cite{Turbide:2003si} of such thermal photon
  emissions with 
  the  pQCD treatment of nucleon-nucleon scattering.
   The high initial temperature
  of the plasma of 270 MeV is lowered to 205 MeV in other scenarios with
  additional $k_{\rm t}$ smearing. Definite conclusions are difficult to draw, due to
  the large background at low $p_{\rm t}$ and the lack of $pPb$ data for a
  direct comparison.
  
  Prompt photons in Au-Au collisions at the higher RHIC energies 
    ($\sqrt{s_{NN}} = 200$ GeV)~\cite{Adler:2005ig}
  are consistent with scaling from $pp$ collisions (Fig.~\ref{fig:nucl}b),
  in remarkable contrast to
  the strongly interacting  $\pi^0$s,
  showing that their suppression in collisions with
   many participating nucleons is a final state effect.

\vspace*{5pt} \hspace{-18pt}
{\bf Acknowledgement :} I am grateful to K. Reygers and J. Turnau for discussions.

\vspace*{50pt}

\begin{table}[htb]
\hspace*{-9pt}
\begin{tabular}{|c|c|c|c|c|c|}
\hline
  \ &reaction&had. energy&d(vtx-calo)&yield/backgd&$R(\eta ,\phi )$ \\
 \hline
 H1/ZEUS & $\gamma p , ep$ & ~200 & 1 &shower analysis& 1 \\
 D0 & $p\bar{p} $ & 1960 & 1 &shower analysis& 0.4 \\
 CDF & $p\bar{p} $ & 1800 & 1 &shower analysis& 0.4 \\
 CDF & $p\bar{p} $ & 1800 & 1 &$\gamma$ conversions& 0.4 \\
 E706 & $pp,pN,\pi N$ & 31, 39& 9 &measure $\gamma/ \pi^0$ & $-$ \\
 WA98 & $PbPb$ & 17.3& 22 &measure $\gamma/ \pi^0$ & $-$ \\
 PHENIX & $AuAu$ & 200& 5 &measure $\gamma/ \pi^0$ & $-$ \\
\hline
 \end{tabular}
\caption{ Characteristic differences of experiments. The hadronic energies
  are in GeV, the distance from vertex to calorimeter in m.
The experiments exploiting explicit $\pi^0$ id, require no isolation
cone $R$.
}
\label{tab1}
\end{table}


\vspace*{-5pt}

\begin{thebibliography}{99}
 
\bibitem{Moses}
Moses, Book 1, 1,2-4.

\bibitem{Breitweg:1999su}
J.~Breitweg {\it et al.}  [ZEUS],
Phys.\ Lett.\ B {\bf 472} (2000) 175
[hep-ex/9910045].

\bibitem{Aktas:2004uv}
A.~Aktas {\it et al.}  [H1],
Eur.\ Phys.\ J.\ C {\bf 38} (2005) 437
[hep-ex/0407018].

\bibitem{Lipatov:2005wk}
A.~V.~Lipatov and N.~P.~Zotov,
hep-ph/0507243;
Phys.\ Rev.\ D {\bf 72} (2005) 054002
[hep-ph/0506044]; 
M.~A.~Kimber, A.~D.~Martin and M.~G.~Ryskin,
Eur.\ Phys.\ J.\ C {\bf 12} (2000) 655
[hep-ph/9911379].



\bibitem{Abbiendi:2003kf}
G.~Abbiendi {\it et al.}  [OPAL],
Eur.\ Phys.\ J.\ C {\bf 31} (2003) 491
[hep-ex/0305075].

\bibitem{Acosta:2002ya}
D.~Acosta {\it et al.}  [CDF],
Phys.\ Rev.\ D {\bf 65} (2002) 112003
[hep-ex/0201004].

\bibitem{D0:2005}
D0note 4859-CONF;http://www-d0.fnal.gov

\bibitem{Apanasevich:2004dr}
L.~Apanasevich {\it et al.}  [Fermilab E706],
Phys.\ Rev.\ D {\bf 70} (2004) 092009
[hep-ex/0407011].

\bibitem{Aggarwal:2000th}
M.~M.~Aggarwal {\it et al.}  [WA98],
Phys.\ Rev.\ Lett.\  {\bf 85} (2000) 3595
[nucl-ex/0006008];
nucl-ex/0006007,
submitted to Phys.Rev.C 


\bibitem{Adler:2005ig}
S.~S.~Adler {\it et al.}  [PHENIX],
Phys.\ Rev.\ Lett.\  {\bf 94} (2005) 232301
[nucl-ex/0503003];
K. Reygers, these proceedings.

\bibitem{Fontannaz:2001ek}
M.~Fontannaz, J.~P.~Guillet and G.~Heinrich,
Eur.\ Phys.\ J.\ C {\bf 21} (2001) 303
[hep-ph/0105121];
Eur.\ Phys.\ J.\ C {\bf 22} (2001) 303
[hep-ph/0107262].
 
\bibitem{Krawczyk:2001tz}
M.~Krawczyk and A.~Zembrzuski,
Phys.\ Rev.\ D {\bf 64} (2001) 114017
[hep-ph/0105166];
 [hep-ph/0309308].

\bibitem{Janssen:2005yz}
X.~Janssen,
these proceedings,
arXiv:hep-ex/0510072.

\bibitem{Chekanov:2004wr}
S.~Chekanov {\it et al.}  [ZEUS],
Phys.\ Lett.\ B {\bf 595} (2004) 86
[hep-ex/0402019].

\bibitem{Aurenche:1983ws}
P.~Aurenche, A.~Douiri, R.~Baier, M.~Fontannaz and D.~Schiff,
Phys.\ Lett.\ B {\bf 140} (1984) 87;   
E.~L.~Berger and J.~w.~Qiu,
Phys.\ Rev.\ D {\bf 44} (1991) 2002.

\bibitem{deFlorian:2005wf}
D.~de Florian and W.~Vogelsang,
Phys.\ Rev.\ D {\bf 72} (2005) 014014
[hep-ph/0506150] and references therein.

\bibitem{Gluck:1994iz}
M.~Gluck, L.~E.~Gordon, E.~Reya and W.~Vogelsang,
Phys.\ Rev.\ Lett.\  {\bf 73} (1994) 388.

\bibitem{Acosta:2004bg}
D.~Acosta {\it et al.}  [CDF],
Phys.\ Rev.\ D {\bf 70} (2004) 074008
[hep-ex/0404022].

\bibitem{Soldner}
S. Soldner-Rembold, these proceedings;
D.~Acosta {\it et al.}  [CDF],
Phys.\ Rev.\ Lett.\  {\bf 95} (2005) 022003
[hep-ex/0412050].

\bibitem{Catani:2002ny}
S.~Catani {\it et al.},
JHEP {\bf 0205} (2002) 028
[hep-ph/0204023].

\bibitem{Turbide:2003si}
S.~Turbide, R.~Rapp and C.~Gale,
Phys.\ Rev.\ C {\bf 69} (2004) 014903
[hep-ph/0308085].


\end{thebibliography}
\end{document}